\documentclass[conference]{IEEEtran}
\IEEEoverridecommandlockouts
\usepackage{cite}
\usepackage{amsmath,amssymb,amsfonts}
\usepackage[ruled,norelsize,vlined,linesnumbered]{algorithm2e}
\usepackage{multirow}
\usepackage{graphicx}
\usepackage{textcomp}
\usepackage{xcolor}
\usepackage{amsmath}  
\usepackage{url}
\usepackage{cite}
\usepackage{hyperref}
\usepackage{siunitx} 
\usepackage{graphicx} 
\usepackage{float}
\usepackage{array}
\usepackage{textcomp}
\usepackage{stfloats}
\usepackage{verbatim}
\usepackage{subfigure}

\usepackage{graphicx}  
\usepackage{caption}   
\usepackage{subcaption} 
\usepackage{cleveref}

\def\BibTeX{{\rm B\kern-.05em{\sc i\kern-.025em b}\kern-.08em
    T\kern-.1667em\lower.7ex\hbox{E}\kern-.125emX}}
\begin{document}
	
\title{IICPilot: An Intelligent Integrated Circuit Backend Design Framework Using Open EDA\\
}

\makeatletter
\newcommand{\linebreakand}{%
\end{@IEEEauthorhalign}
\hfill\mbox{}\par
\mbox{}\hfill\begin{@IEEEauthorhalign}Backend
}
\makeatother

\author{
	\IEEEauthorblockN{Zesong Jiang$^{1,2}$, Qing Zhang$^{1}$, Cheng Liu$^{1,3}$\IEEEauthorrefmark{1}\thanks{\IEEEauthorrefmark{1} Corresponding author.}, Long Cheng$^{4}$, Huawei Li$^{1,3}$, Xiaowei Li$^{1,3}$}
	\IEEEauthorblockA{
		$^{1}$SKLP, Institute of Computing Technology, Chinese Academy of Sciences, Beijing, China
	}
	\IEEEauthorblockA{
		$^{2}$Institute of Advanced Technology, University of Science and Technology of China, Hefei, China
	}
	\IEEEauthorblockA{
		$^{3}$Dept. of Computer Science, University of Chinese Academy of Sciences, Beijing, China
	}
       \IEEEauthorblockA{
		$^{4}$School of Control and Computer Engineering, North China Electric Power University, Beijing, China
	}
	\vspace{-2em}

 \thanks{This work is supported by the National Key R\&D Program of China under Grant (2022YFB4500405), and the National Natural Science Foundation of China under Grant 62174162.}
}

\maketitle
\begin{abstract}
    Open-source EDA tools are rapidly advancing, fostering collaboration, innovation, and knowledge sharing within the EDA community. However, the growing complexity of these tools, characterized by numerous design parameters and heuristics, poses a significant barrier to their widespread adoption. This complexity is particularly pronounced in integrated circuit (IC) backend designs, which place substantial demands on engineers' expertise in EDA tools. To tackle this challenge, we introduce IICPilot, an intelligent IC backend design system based on LLM technology. IICPilot automates various backend design procedures, including script generation, EDA tool invocation, design space exploration of EDA parameters, container-based computing resource allocation, and exception management. By automating these tasks, IICPilot significantly lowers the barrier to entry for open-source EDA tools. Specifically, IICPilot utilizes LangChain's multi-agent framework to efficiently handle distinct design tasks, enabling flexible enhancements independently. Moreover, IICPilot separates the backend design workflow from specific open-source EDA tools through a unified EDA calling interface. This approach allows seamless integration with different open-source EDA tools like OpenROAD and iEDA, streamlining the backend design and optimization across the EDA tools.
\end{abstract}


\IEEEpeerreviewmaketitle

\section{Introduction}


Electronic Design Automation (EDA) occupies an important position in chip design and affects the PPA (performance, power, and area) of the resulting designs substantially \cite{chen2024dawn}. As a critical interface between the chip design and fabrication, it converts the gate-level netlist generated in front-end to manufacturable GDSII data. However, with the continuous increase in chip design complexity and the accelerated pace of technological innovation, the backend design of integrated circuits faces multiple-folded challenges. On one hand, the backend design of chips involves a set of complex design procedures that require the use of various EDA software and design tools, each with complex programming interfaces and data formats, making automated chip backend design highly challenging. On the other hand, advanced backend design tools typically contain a large number of adjustable design parameters, and exploring the design space across a large number of options remains a pressing issue to be resolved. In addition, unlike commercial EDA tools, open-source EDA tools such as iEDA\cite{li2023ieda} and OpenROAD \cite{ajayi2019openroad} are usually less robust due to the limited manpower and financial support, and unfamiliar to most of the designers, which further discourages the use of the open EDA tools.

Hence, we argue that automating the use of Open EDA tools can substantially mitigate the barrier of using these EDA tools and encourage more feedback for continuous improvement. Although there are already many scripts developed to invoke these EDA tools conveniently, it remains insufficient for automating the backend design as it still require users to tune the configurations and parameters to suit the different designs and constraints. Motivated by the successful adoption of large language models (LLMs) on more and more complex tasks such as robotics and autonomous drving, we also attempt to leverage the powerful reasoning and natural language understanding capabilities of LLM to fully automate the use of Open EDA tools and make it accessible to more designers without backend design expertise. 
In this context, we propose, IICPilot, an LLM-based automatic IC backend design framework using open EDA. Essentially, it is a multi-agent system based on LLMs and has each agent specialized for a relatively independent task such as floorplan and routing such that each agent can focus on a relatively short context and be updated without affecting the other agents. Particularly, it has a user proxy agent to understand the user requirements through multiple-round interaction using natural language. Then, it has a control agent to leverage chain-of-thoughts and automatically produce a task sequence based on user requirements. Hence, the framework can be adapted to various backend designs. For instance, it can conduct design space exploration of the entire backend design flow or a specific backend design process. Basically, IICPilot can autonomously generate scripts, execute EDA tasks, and optimize chip PPA through design space exploration tools and mitigate the barrier of using open EDA tools. Additionally, it can allocate appropriate computing resources through containers to sustain various complex EDA tasks which may involve a set of time-consuming yet dependent EDA procedures. This not only effectively break the barrier of using open EDAs but also scales the complex backend design tasks over a distributed computing system for higher performance.

The major contributions of this work are as follows:
\begin{itemize}
    \item We introduce IICPilot, the first intelligent backend design framework offering full-stack automation including backend design and distributed deployment for open EDA. 
\end{itemize}
\begin{itemize}
    \item we propose a container agent that can automatically allocate appropriate computing resources for various EDA tasks and RTL designs, and provide scalable runtime optimization on distributed computing systems.
\end{itemize}
\begin{itemize}
    \item We propose a DSE agent that can automatically extract and adjust the parameters of open EDA tools for the sake of better PPA.
\end{itemize}

\section{Background \& Related Work}

\subsection{LLM-based Design Automation}
LLMs such as GPT-3.5 and GPT-4 have achieved significant milestones in the field of natural language processing, offering robust support for research and applications with their exceptional performance and extensive applications. Multi-agent systems, comprised of collaborative agents, possess the ability to independently perceive, make decisions, and interact with each other. Following the advent of cutting-edge LLMs, the utilization of intelligent agents has been propelled into a new era of prominence and significance.

Recently, multi-agent systems based on LLMs have been applied across various fields, including software design\cite{hong2023metagpt, qian2023communicative, wu2023autogen, nakajima2023babyagi}, robotics\cite{mandi2023roco, zhang2023building}, social simulation\cite{park2023generative}, and game simulation\cite{xu2023language}, significantly enhancing project efficiency and simulation outcomes. 
Concurrently, there is also significant research on LLMs in the context of IC backend development. HDL debugger\cite{yao2024hdldebugger} uses large models and RAG technology to debug hardware description languages. VeriGen\cite{thakur2023benchmarking} has improved by scaling up the model size and expanding the hardware dataset. RTLLM\cite{lu2024rtllm} and VerilogEval\cite{liu2023verilogeval} have introduced larger-scale open benchmarks for designing RTL generation based on natural language, evaluating prompt and fine-tuned models on these benchmarks. Chip-Chat\cite{blocklove2023chip} aims to assess the collaborative effectiveness of GPT-4 with hardware designers in generating processors and completing tape-outs. ChatEDA \cite{wu2024chateda} automates EDA tools using LLMs. RTLCoder \cite{liu2023rtlcoder}, CodeGen \cite{nijkamp2022codegen}, VeriAssist \cite{huang2024towards}, and AutoChip \cite{thakur2023autochip} achieve RTL code generation through LLMs. More studies on LLMs for EDA can be referenced in this survey \cite{zhong2023llm4eda}.

Despite the abundance of related research \cite{chen2024dawn}, the integration of LLMs with multi-agent systems for intelligent work in IC backend design remains unexplored. By integrating LLMs and multi-agent systems, we can create a more intelligent and efficient dedicated IC backend system that can deeply analyze user needs and automate complex tasks, thus alleviating the burden on engineers and enhancing design efficiency and quality.

\subsection{Design Space Exploration of CAD Tools}

In IC backend design, Design Space Exploration (DSE) systematically evaluates various design options and parameter combinations to determine the optimal design that meets specific performance, power consumption, area, and cost requirements.

The application of DSE offers significant advantages: it enables designers to quickly identify the optimal solution within a complex design space, eliminating the need for blind trials and lengthy iterations characteristic of traditional design processes. Additionally, DSE provides a range of alternative design options, allowing designers to select the most suitable design based on their specific needs.

With the ongoing evolution of artificial intelligence and machine learning technologies, many efforts\cite{bai2021boom,ma2019cad,geng2022ppatuner,geng2022techniques,geng2022ptpt,zhai2023microarchitecture} have significantly boosted chip performance by integrating these advanced techniques into DSE. Currently, DSE plays a crucial role in IC backend design, driving the continuous advancement of the integrated circuit design field. This article aims to leverage open-source tools to complete the DSE and achieve the best combination of backend parameters.

\subsection{Kubernetes and Containers}
Kubernetes (K8s) is a powerful open-source container orchestration system designed for automating container deployment, scaling, and management. As a lightweight virtualization technology, containers encapsulate applications and their dependencies into portable units, enabling seamless migration and operation across different environments.


In our research, the automation capabilities of K8s are crucial. It automatically handles container deployment, scaling, scheduling, and fault recovery, ensuring optimal task performance within the cluster. Futhermore, the intelligent agent in our system can allocates and adjusts resources based on the specific needs of IC backend tasks through K8s, maximizing design throughput and efficiency. These features make K8s an ideal choice for managing IC backend operations.
\begin{figure*}
    \centering
    \setlength{\abovecaptionskip}{0.2cm} 
    \setlength{\abovecaptionskip}{0.cm} 
    \includegraphics[width=0.9\linewidth,height=9cm]{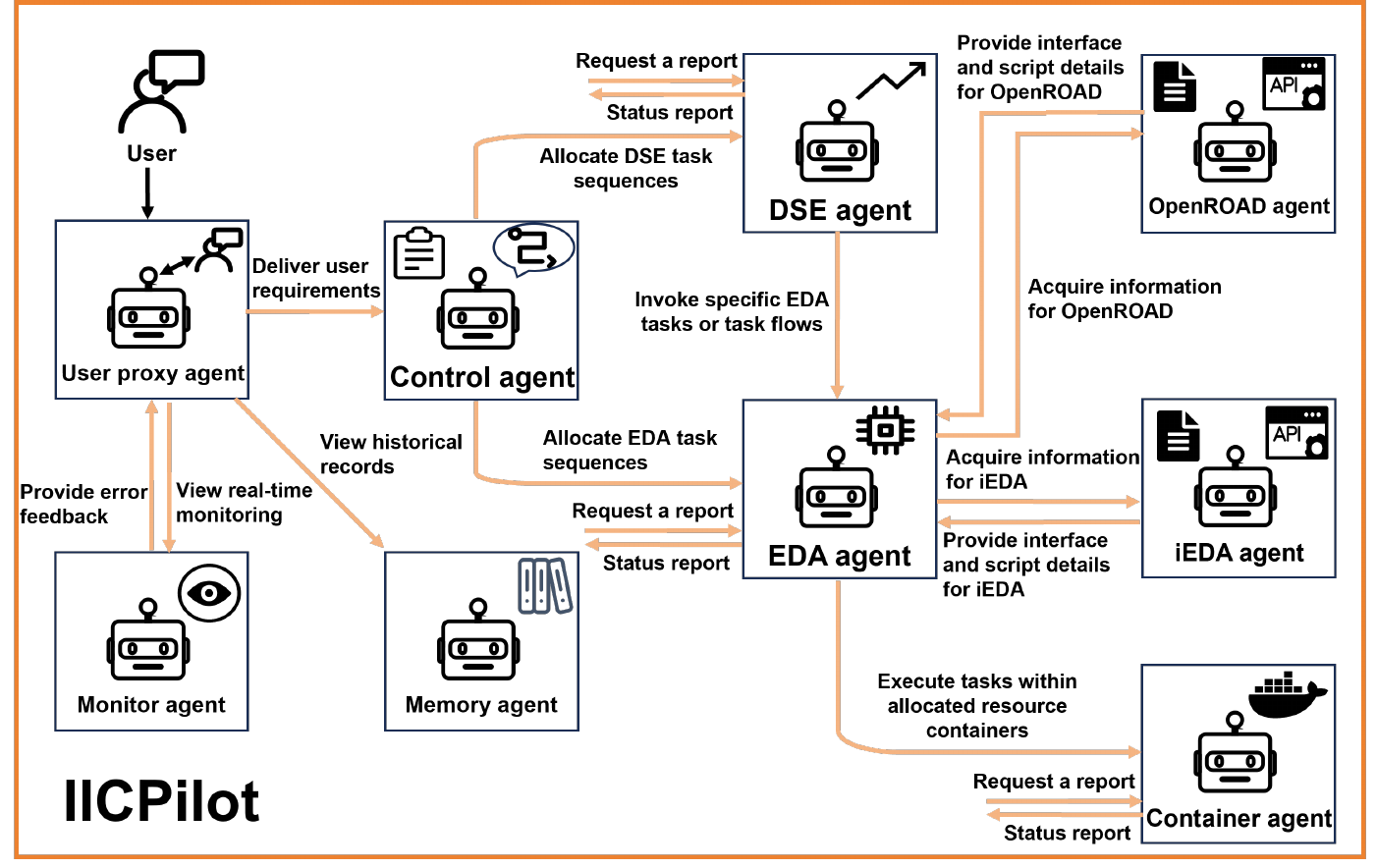}
    \caption{IICPilot Overview.}
    \label{construction}
    
\end{figure*}
\section{IICPilot Framework}
In this study, we introduce IICPilot, a multi-agent driven end-to-end EDA optimization framework. The aim of this framework is to implementing intelligent solutions for EDA tasks. IICPilot can automate IC backend tasks through collaborative efforts of multiple agents. Unlike the previous LLM used in EDA tasks, this framework not only addresses script generation for running EDA tasks, but also enhances chip performance using DSE tools and considers resource savings through containerization. It proposes solutions for runtime failures in practical scenarios and importantly, supports compatibility with two open-source platforms, demonstrating practical utility. Additionally, the framework can real-time monitor agent operations, obtain status reports, and access historical records of executions.

\subsection{LLM-based Agent Construction}
As illustrated in \Cref{construction},for each agent within the IICPilot system, we have meticulously designed their architecture, selected and developed various tools, and ensured that the agents can effectively utilize these tools.


Firstly, the user agent acts as the interface between the entire framework and the user, responsible for translating user requirements into feasible IC backend tasks. It can also gather information from users when deemed insufficient by other agents. The control agent, on the other hand, transforms user requirements into a task list and assigning tasks to other agents. EDA agent generates or modifies the EDA task scripts according to the task list and executes different backend design processes. To enhance its generation capability for various EDA tasks, we have also equipped the EDA agent with the ability to use tools and understand EDA documentation.

Furthermore, we observed that the agent's processing and understanding of large volumes of information can sometimes be confused. To address this, we have classified the information in EDA tasks into essential and optional categories, aiming for the agent to extract effective information before completing the task. Essential information refers to the data necessary to complete the EDA task, while optional information encompasses additional optimization details provided by the user to improve task outcomes. In this system, essential information includes RTL design and the specific EDA stage, the selection of technology node, and constraint files. Optional information includes clock period, core area utilization, placement density, etc. Essential information generally has well-defined paths and details, and we only need to articulate the requirements. However, optional information involves multiple dimensions and requires users to provide file paths, filenames, and specific modification values to facilitate task completion. Figure \ref{example_detail} illustrates an example about how EDA agent works based on this method.
\begin{figure}[ht]
    \vspace{-0.0cm}
    \centering
    \includegraphics[width=0.9\linewidth,height=8cm]{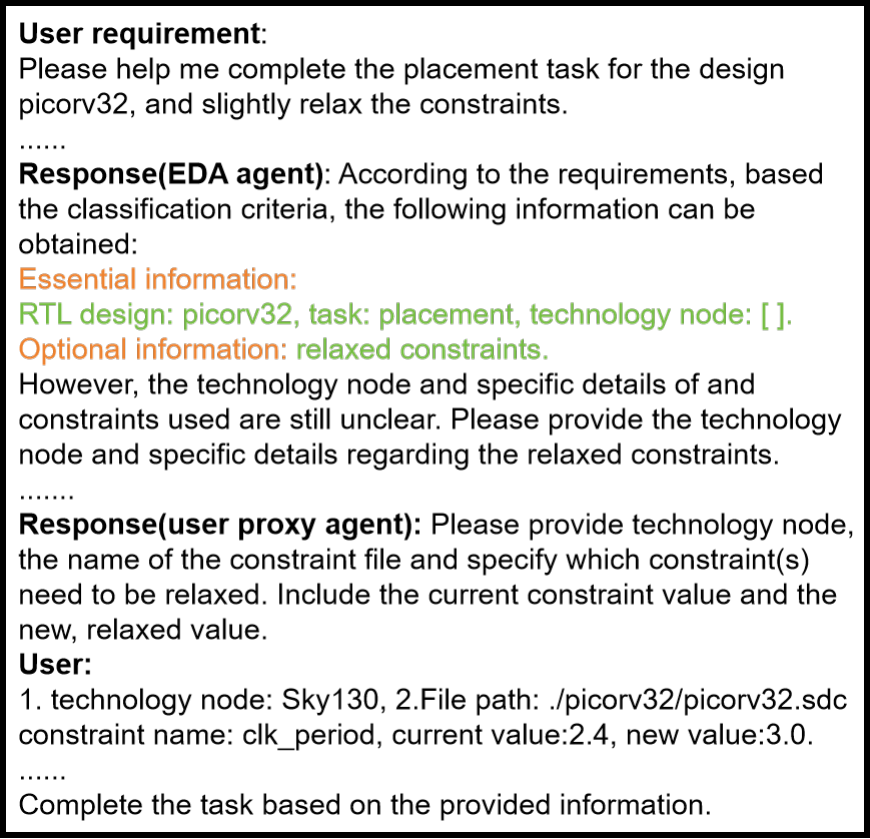}
    \caption{An example explains how an agent executes the task based on essential and optional information.}
    \label{example_detail}
    \vspace{-0.4cm}
\end{figure}

The DSE agent conducts design space exploration to optimize IC backend configuration parameters, enhancing chip performance. We define the role and objectives for the DSE agent and provide it with tools for modifying JSON-formatted parameter configuration files. To facilitate the process of design space exploration, we equip the agent with design space exploration tools (Autotuner\cite{jung2021metrics2} and Hypermapper\cite{nardi2019practical}) and DSE agent will invoke the EDA agent to complete EDA tasks during iterations through dse tools. Additionally, To address potential issues where unreasonable parameter ranges hinder exploration, DSE agent can consult the fault list to find solutions. This fault list will be introduced in the \ref{se-workflow} section.

Furthermore, for enhanced compatibility across two platforms, we have developed the iEDA agent and OpenROAD agent. Each agent holds comprehensive resources pertaining to iEDA and OpenROAD respectively. When EDA agent needs to accomplish a task using a specific open-source toolset, it can request specific information and interfaces from these agents, thereby facilitating seamless execution of EDA tasks.

The container agent create a container with suitable machine resources for EDA or DSE tasks, optimizing resource utilization and reducing costs. To achieve this, we provided it with K8s API for flexible container creation and resource allocation. Since machine resource cost is calculated as the product of the cost per unit time of different machine configurations and the runtime, we aim to predict the time required to complete specific tasks under various machine configurations. In general, the runtime of EDA tasks correlates with the scale of the RTL design, computational resources, and the nature of the tasks involved. To address this, we employ machine learning techniques to forecast EDA task runtimes across various stages, considering the RTL design size and machine configuration. The container agent can invoke trained model to allocate resources efficiently for a specific design and EDA stage. 
\begin{figure*}[htbp]
	\centering
        \vspace{-1.5cm}   
        \setlength{\abovecaptionskip}{0.cm} 
        \setlength{\abovecaptionskip}{0.cm} 
        \setlength{\belowdisplayskip}{3pt}
	\subfigure[]{
		\begin{minipage}[t]{0.5\linewidth}
			\centering
			\includegraphics[width=1.0\linewidth,height=7cm]{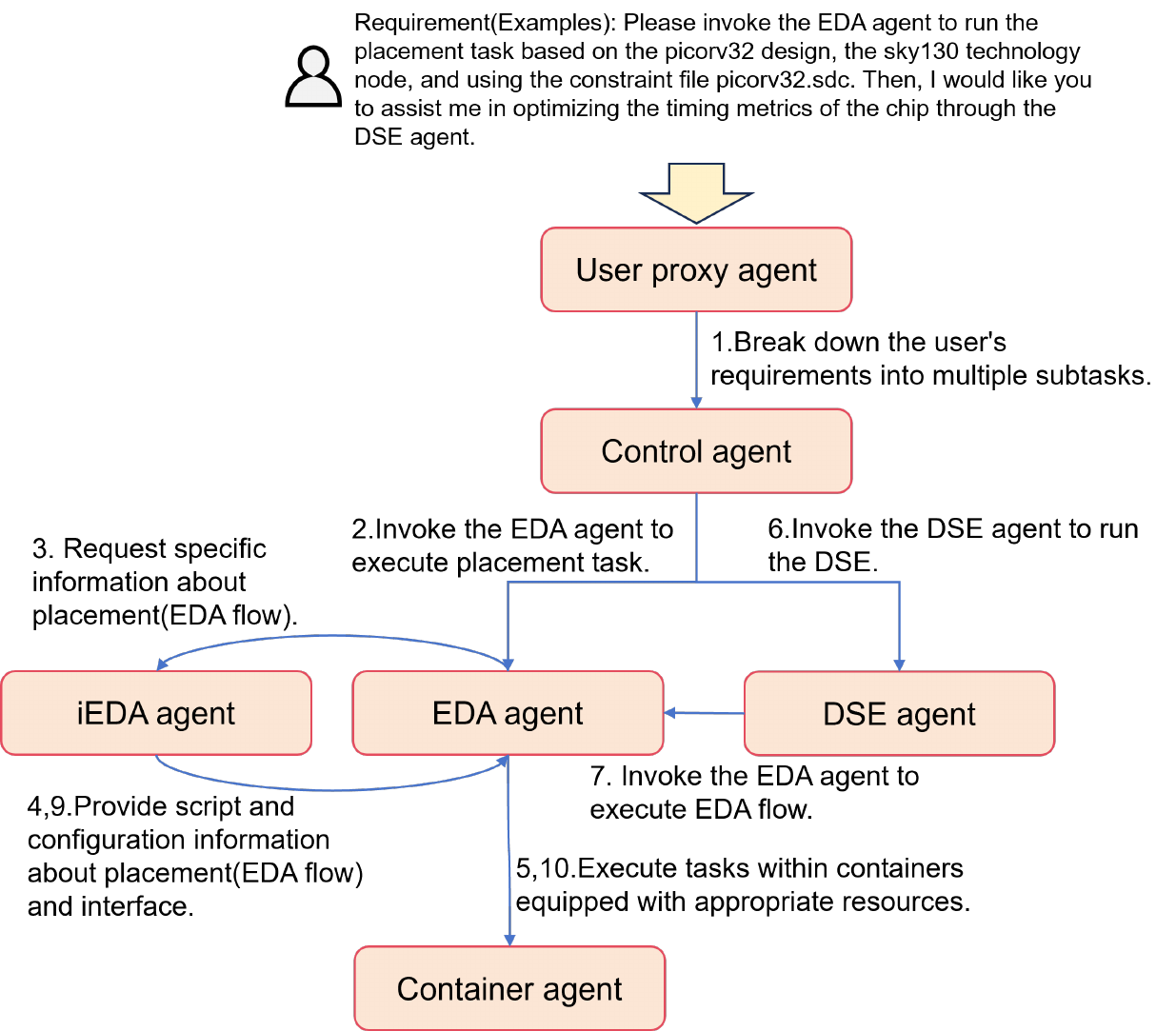}\\
			\vspace{0.02cm}
		\end{minipage}%
            \label{ex1}
	}%
	\subfigure[]{
		\begin{minipage}[t]{0.45\linewidth}
			\centering
			\includegraphics[width=1.0\linewidth,height=7cm]{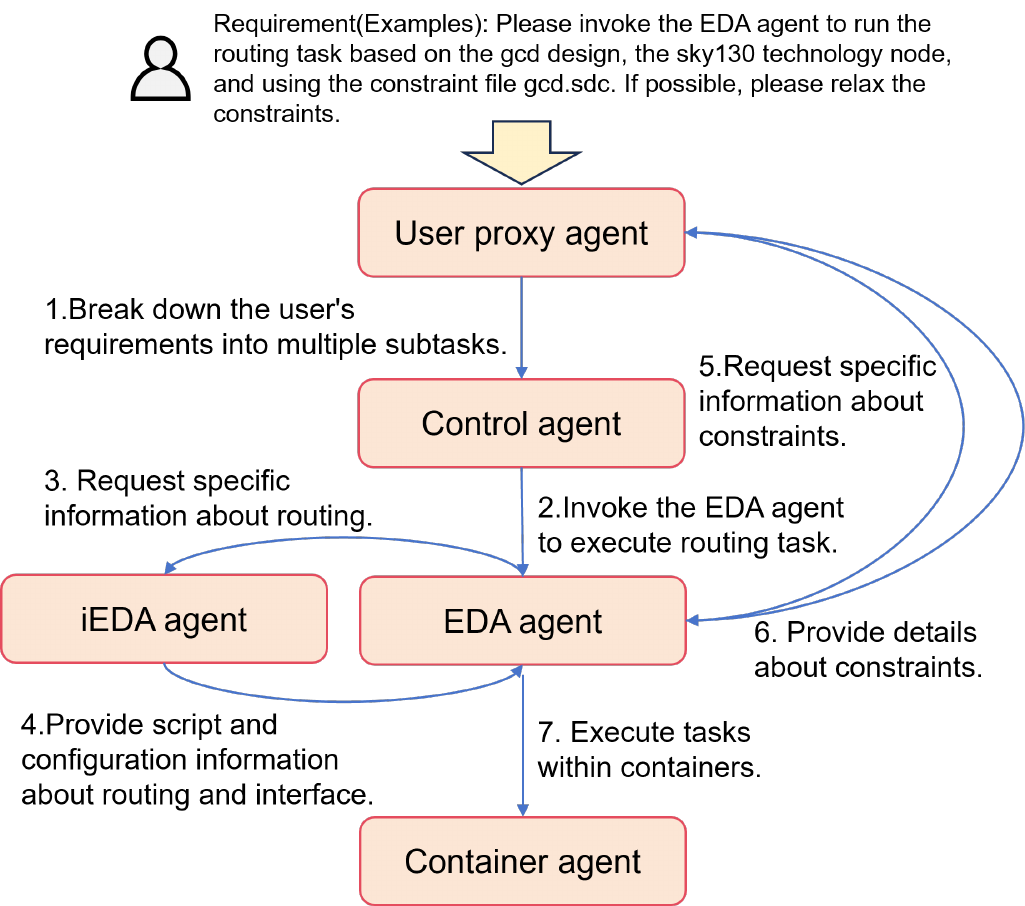}\\
			\vspace{0.02cm}
		\end{minipage}%
            \label{ex2}
	}%
	\centering
	\caption{Examples of using IICPilot.}
	\vspace{-0.0cm}
	\label{ex}
\end{figure*}
Furthermore, for EDA problems involving multiple stages or entire flows with time constraints (a scenario of significant practical value, given that most chip design-to-tapeout processes are not only constrained by resources but also have deadlines for completion), we adapt the approach from \cite{hosny2021characterizing}, \cite{kellerer2004multiple}, mapping the problem to the Multiple-Choice Knapsack Problem(MCKP) and obtain a solution. Specifically, each EDA subtask can be run on different machine configurations (i.e., vCPUs), which complete the job in t time and costs p.
Let $y_{m}(C)$ be an optimal solution defined on m applications and with total time constraints C:
\begin{equation}
    y_{m}(C):=max \sum_{i=1}^{m}  \sum_{j=1}^{K_{i}} s_{ij} \frac{1}{c_{ij}}
\end{equation}
such that,
\begin{equation*}
        \sum_{i=1}^{m}  \sum_{j=1}^{K_{i}} s_{ij} \frac{1}{c_{ij}} \leq C
\end{equation*}
\begin{equation*}
        \sum_{j \in K_i} s_{ij} = 1, \quad i=1,\ldots,m
\end{equation*}
\begin{equation*}
       s_{ij} \in \{0, 1\}, \quad i=1,\ldots,m, \quad j \in K_i
\end{equation*}
where $s_{ij}$ denotes whether we select configuration $j$ for subtasks $i$ or not, and $K_i$ is the number of configurations used in the subtask. Similarly, $c_{ij}$ represents the expenditure incurred for executing stage $i$ with configuration $j$, and we get this from the price list of the designated cloud service provider. To find the optimal configuration, we implemented a pseudopolynomial solution using dynamic programming based on the method from Dudzinski and Walukiewicz \cite{dudzinski1987exact}.

\begin{equation*}
y_m(C) = \max \left\{
\begin{array}{ll}
    y_{m-1}(C - t_{m1}) + {1}/{c_{m1}} & \text{if } 0 \leq C - t_{m1}, \\
    y_{m-1}(C - t_{m2}) + {1}/{c_{m2}} & \text{if } 0 \leq C - t_{m2}, \\
    \vdots & \\
    y_{m-1}(C - t_{mn_l}) + {1}/{c_{m_l}} & \text{if } 0 \leq C - t_{m_l}.
\end{array}
\right.
\end{equation*}

Using this strategy, we can determine the optimal CPU configuration selection for multiple EDA tasks with time constraints.


Considering practical application scenarios, we have also designed the monitor agent and memory agent. The monitor agent monitors the real-time operation of agents, providing timely feedback to users in case of issues. Additionally, users can request status reports from specific agents to understand their operational status. The memory agent maintains system operation records, allowing users to access historical data.



\begin{figure}
    \vspace{-0.1cm}
    \centering
    \includegraphics[width=0.9\linewidth,height=6.5cm]{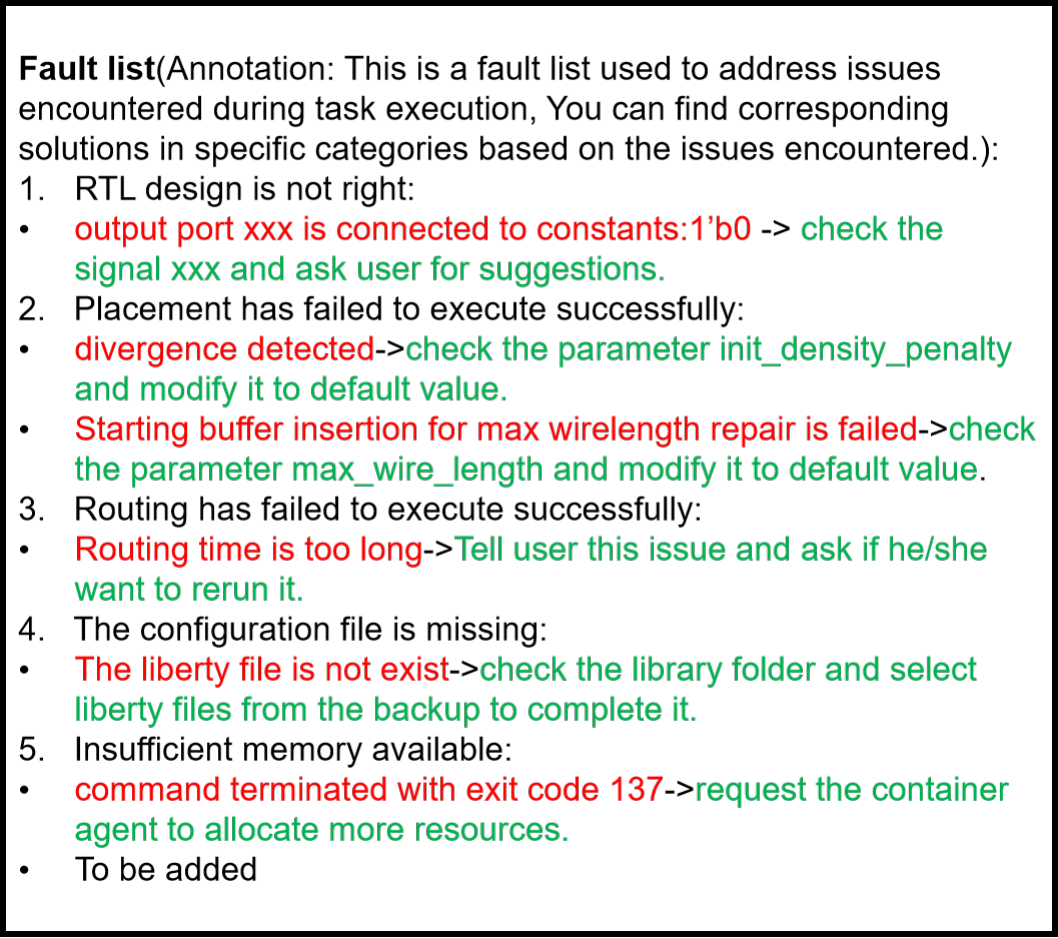}
    \caption{The fault list used for self-correction..}
    \label{fault_list}
    \vspace{-0.4cm}
\end{figure}
\begin{table*}[h]
  \centering
  \vspace{-0.5cm}
  \scalebox{1.1}{
    \begin{tabular}{|c|c|ccc|ccc|c|}
    \hline
    \multirow{2}{*}{Design} & \multirow{2}{*}{clk\_period} & \multicolumn{3}{c|}{Default}                                                       & \multicolumn{3}{c|}{IICPilot}                                                      & \multirow{2}{*}{Optimization} \\ \cline{3-8}
                            &                              & \multicolumn{1}{c|}{cp\_delay/(ns)} & \multicolumn{1}{c|}{power/(mw)} & area/(um²) & \multicolumn{1}{c|}{cp\_delay/(ns)} & \multicolumn{1}{c|}{power/(mw)} & area/(um²) &                               \\ \hline
    aes                     & 0.82                         & \multicolumn{1}{c|}{1.1529}         & \multicolumn{1}{c|}{447.0}      & 53141      & \multicolumn{1}{c|}{1.1360}         & \multicolumn{1}{c|}{418.0}      & 53200      & 7.76\%                        \\ \hline
    picorv32                & 2.5                          & \multicolumn{1}{c|}{1.9330}         & \multicolumn{1}{c|}{24.8}       & 57593      & \multicolumn{1}{c|}{1.9304}         & \multicolumn{1}{c|}{24.0}       & 40079      & 32.75\%                       \\ \hline
    ibex                    & 2.8                          & \multicolumn{1}{c|}{2.2622}         & \multicolumn{1}{c|}{102.0}      & 57987      & \multicolumn{1}{c|}{2.2569}         & \multicolumn{1}{c|}{102.0}      & 48305      & 16.89\%                       \\ \hline
    gcd                     & 0.46                         & \multicolumn{1}{c|}{0.4086}         & \multicolumn{1}{c|}{2.9}        & 1143       & \multicolumn{1}{c|}{0.4136}         & \multicolumn{1}{c|}{3.1}        & 960        & 9.12\%                        \\ \hline
    \end{tabular}
  }
  \caption{The automated execution of DSE tasks by the IICPilot.(Optimization refers to the percentage increase in the product of PPA performance after DSE compared to that before DSE, cp\_delay refers to the critical path delay, and the technology used in this experiment is nangate45.)}
  \label{e3}
\end{table*}
\subsection{The Workflow of IICPilot}
\label{se-workflow}

As illustrated in \Cref{ex}, we provide two descriptive examples to elaborate on the system's workflow. In the context of \Cref{ex1}, our multi-agent system initially captures user requirements through the User proxy agent and forwards them to the control agent, which decomposes these requirements into a series of subtasks executable by intelligent agents. Specifically, EDA tasks are initially assigned to the EDA agent. Upon receiving such tasks, the agent first retrieves process-specific information from the corresponding platform agent (typically the iEDA agent). EDA agent executes the required EDA tasks accordingly. Following this are DSE tasks, where the DSE agent is activated to generate or modify parameter configuration files and iterate through the EDA process via the EDA agent. It is noteworthy that to optimize machine resource utilization and minimize task completion costs, these tasks are executed within containers. Specifically, upon completion of each subtask, commands and configuration files seamlessly transfer to the container agent, which determines the machine resources required for container allocation. This comprehensive process covers the entire automated workflow, from user input to IC backend task generation, culminating in execution at the container level. For \Cref{ex2}, it reflects that if the user does not provide specific constraints or detailed information, the agent will proactively request additional information from the user.

Furthermore, we provide a fault list where, upon encountering errors during execution. We expect the internal agents of the system to refer to this documentation to autonomously identify solutions and self-correct based on its contents, primarily focusing on the DSE agent, as DSE involves iterative optimization, numerous parameter-related issues may arise during the process, making it relatively complex. Users can contribute to this documentation by adding encountered errors and their corresponding solutions, gradually achieving comprehensive coverage of error scenarios. We have found that agents can effectively address issues in the figure \Cref{fault_list} with the help of the list itself.

\section{Experiment}
\begin{figure}
    \vspace{-0.5cm}
    \centering
    \includegraphics[width=0.8\linewidth,height=7cm]{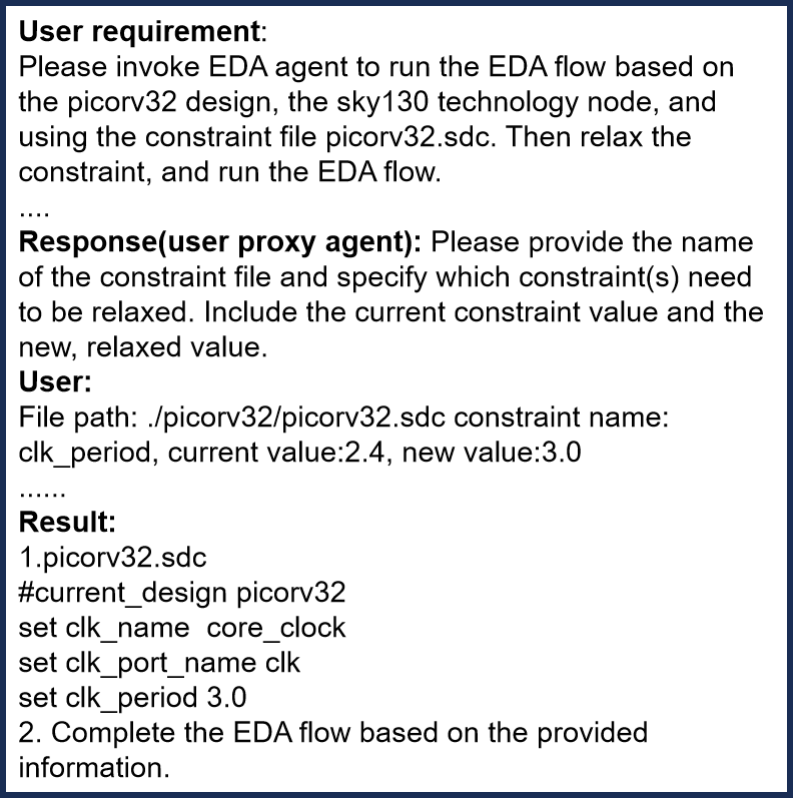}
    \caption{Automated execution of EDA tasks generated by IICPilot.}
    \label{e2}
    \vspace{-0.3cm}
\end{figure}
In this section, we evaluate the IICPilot framework, examining its capabilities ranging from automated execution of EDA tasks to optimizing chip performance and reducing machine resource costs. Futhermore, we delve into the key techniques employed in IICPilot and substantiate their benefits through thorough experimental validation.
\subsection{Experiment Setup}
To evaluate the effectiveness of the IICPilot framework, we conducted a series of experiments using open-source EDA tools. Prior to initiating the experimental phase, we selected iEDA and OpenROAD as our platforms and utilized Autotuner and Hypermapper for design space exploration. Additionally, we configured an appropriate K8s environment and deployed four nodes on Alibaba Cloud. We also gathered 400 data points from open-source websites such as Opencores\cite{opencores} to complete the container-related experiments.

\subsection{Evaluation on EDA Tasks}
    
    In the first set of experiments, we aim to evaluate the efficacy of multi-agent systems in executing EDA tasks. Due to the length of the article, we will only demonstrate the effectiveness of the tool on iEDA. Without the need to invoke the planning agent, tasks can be smoothly completed relying on the EDA agent and the user agent. As shown in \Cref{e2}, an exemplary experiment demonstrates the successful completion of corresponding tasks executed by the EDA agent and user proxy agent.

\begin{table*}[htbp]
  \centering
  \vspace{-0.5cm}
  \renewcommand{\arraystretch}{1.5} 
  \scalebox{1}{ 
    \begin{tabular}{|c|cccc|cccc|cccc|c|}
\hline
Task          & \multicolumn{4}{c|}{placement}                                               & \multicolumn{4}{c|}{routing}                                                 & \multicolumn{4}{c|}{STA}                                                     & Best combination \\ \hline
vCPUs   & \multicolumn{1}{c|}{1} & \multicolumn{1}{c|}{2} & \multicolumn{1}{c|}{4} & 8 & \multicolumn{1}{c|}{1} & \multicolumn{1}{c|}{2} & \multicolumn{1}{c|}{4} & 8 & \multicolumn{1}{c|}{1} & \multicolumn{1}{c|}{2} & \multicolumn{1}{c|}{4} & 8 &     (4,8,1)          \\ \hline

runtime(sec.)  & \multicolumn{1}{c|}{346}  & \multicolumn{1}{c|}{172}  & \multicolumn{1}{c|}{70}  &  65 & \multicolumn{1}{c|}{1966}  & \multicolumn{1}{c|}{1110}  & \multicolumn{1}{c|}{414}  & 378 & \multicolumn{1}{c|}{19}  & \multicolumn{1}{c|}{16}  & \multicolumn{1}{c|}{14}  & 12  &         467         \\ \hline

cost(CNY.)   & \multicolumn{1}{c|}{3.728}  & \multicolumn{1}{c|}{3.471}  & \multicolumn{1}{c|}{2.176}  &  3.282 & \multicolumn{1}{c|}{21.193}  & \multicolumn{1}{c|}{22.418}  & \multicolumn{1}{c|}{12.880}  &  19.082 & \multicolumn{1}{c|}{0.205}  & \multicolumn{1}{c|}{0.323}  & \multicolumn{1}{c|}{0.435}  & 0.606  &     21.465          \\ \hline
\end{tabular}
  }
  \caption{The automated execution of resources allocation based on multi-task by the IICPilot.(Complete the backend design flow for picorv32 with a timing constraint of 480 seconds.}
  \label{e5}
\end{table*}

\begin{table}[htbp]
\centering
\vspace{-0.0cm}
\scalebox{1.0}{
    \begin{tabular}{|c|cccc|}
    \hline
    Task          & \multicolumn{4}{c|}{Placement(picorv32)} \\ \hline
    vCPUs         & \multicolumn{1}{c|}{1}      & \multicolumn{1}{c|}{2}      & \multicolumn{1}{c|}{4}       & 8       \\ \hline
    runtime(sec.) & \multicolumn{1}{c|}{346}    & \multicolumn{1}{c|}{172}    & \multicolumn{1}{c|}{70}      & 65      \\ \hline
    cost/h(CNY.)    & \multicolumn{1}{c|}{38.790} & \multicolumn{1}{c|}{72.650} & \multicolumn{1}{c|}{111.920} & 181.750 \\ \hline
    total cost(CNY.)& \multicolumn{1}{c|}{3.728}  & \multicolumn{1}{c|}{3.471}  & \multicolumn{1}{c|}{2.176}   & 3.282   \\ \hline
    \end{tabular}}
    \caption{The automated execution of resources allocation based on single-task by the IICPilot.}
    \label{e4}
\end{table}

\subsection{Evaluation on DSE tasks}

In the second experimental group, we aim to evaluate the capabilities of a multi-agent system in executing DSE tasks. We conduct multiple experiments using the Autotuner from OpenROAD, validating the system's effectiveness through empirical results. As illustrated in \Cref{e3}, the experiments explore the backend parameter spaces of picorv32, ibex, gcd, and aes to assess the system's impact on chip performance optimization. The multi-agent system is tasked with utilizing DSE agents to complete the design space exploration for specific designs, focusing on optimizing metrics such as area, power consumption, and critical path delay. The results demonstrate that the multi-agent system effectively improves the performance metrics across different designs. 
\subsection{Evaluation on Resources Allocation}
In this experiment, firstly, to illustrate the advantages of containerization and multi-node deployment over non-deployment scenarios, we deployed four nodes in the cloud. We conducted multiple experiments running eight different RTL designs on configurations utilizing 1, 2, 3, and 4 nodes respectively. As shown in \Cref{cloud}, compared to non-deployed environments, the multi-node container deployment architecture based on Kubernetes enables concurrent execution of multiple tasks, demonstrating significant speed advantages. Then, we aim to test the capability of a multi-agent system using containers in intelligent resource allocation. To achieve this, we take configuration as an example and verify the system's ability through experimental results. In this experiment, we first complete the EDA subflow for 400 benchmarks from open-source websites (e.g., OpenCores) under different settings (1vCPU, 2vCPUs, 4vCPUs, 8vCPUs) to measure the runtime under various configurations. Based on these measurements, a simple dataset is created, and a random forest model is employed for prediction, utilizing features such as the number of cells that reflect the design size and machine configuration. The container agent can invoke this trained model to predict the runtime of a specific design under a particular configuration. Then, the cost of each configuration for eash EDA stage is calculated based on the unit-hour running cost of different CPU configurations on Alibaba Cloud, resulting in the lowest-cost machine configuration. Finally, container agent will allocate corresponding resources to the container and completes the EDA tasks within the container based on this result. \Cref{e4} provides an example of using machine learning to predict runtime and obtain containers at minimum cost. We can observe that when user need to accomplish the placement task for the rtl design picorv32, it is appropriate for the CPU to select four units. In addition, \Cref{e6} illustrates the accuracy of the system's invocation model in predicting time. Furthermore, for EDA problems involving multiple subtasks or entire flows with time constraints, as depicted in \Cref{e5}, container agent will give a solution based on the MCKP problem to achieve optimal resource allocation within the system.
\begin{figure}
    \vspace{-0.4cm}
    \centering
    \includegraphics[width=0.8\linewidth,height=5.5cm]{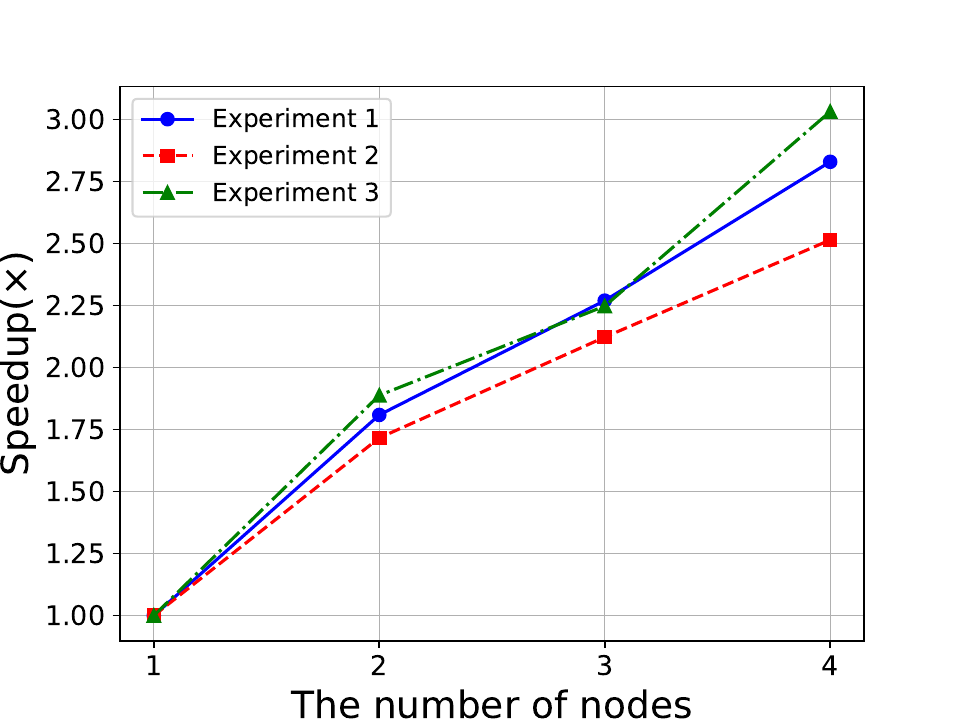}
    \caption{The speedup ratio of executing EDA tasks using multiple nodes.}
    \label{cloud}
    \vspace{-0.0cm}
\end{figure}

\begin{figure}
    \vspace{-0.4cm}
    \centering
    \includegraphics[width=0.8\linewidth,height=5.5cm]{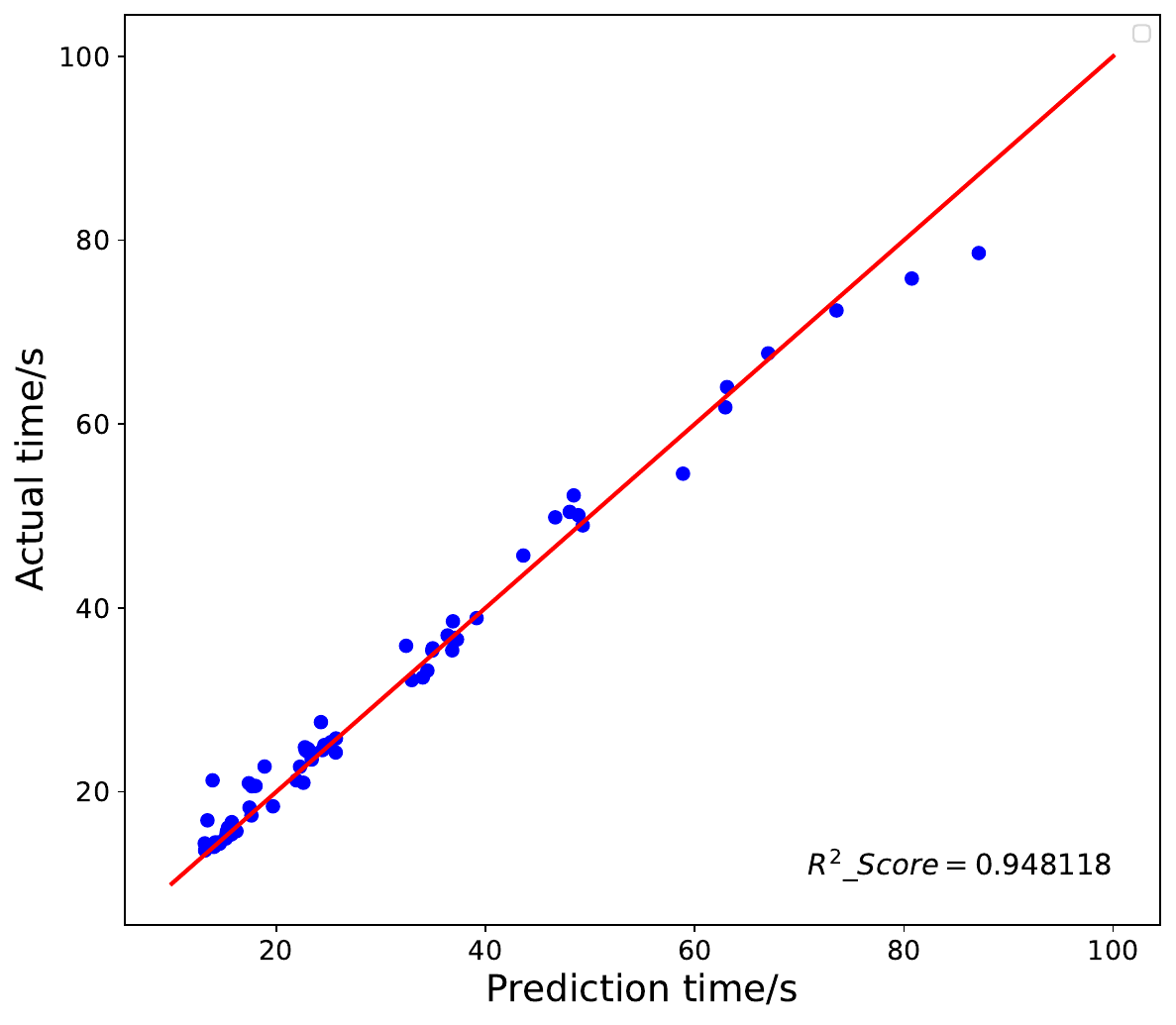}
    \caption{The accuracy in predicting time for running placement of various designs.}
    \label{e6}
    \vspace{-0.0cm}
\end{figure}

\section{Conclusion}
This paper realizes a multi-agent system dedicated to IC backend to complete various tasks in IC backend, which can assist engineers' work and effectively save their energy. The efficiency of this system has been effectively verified based on the open-source tools iEDA and OpenROAD. It is believed that through continuous optimization, this system will play a greater role in IC backend work in the future.

\newpage
\bibliographystyle{IEEEtran}
\bibliography{IICPilot}

\begin{thebibliography}{10}
\providecommand{\url}[1]{#1}
\csname url@samestyle\endcsname
\providecommand{\newblock}{\relax}
\providecommand{\bibinfo}[2]{#2}
\providecommand{\BIBentrySTDinterwordspacing}{\spaceskip=0pt\relax}
\providecommand{\BIBentryALTinterwordstretchfactor}{4}
\providecommand{\BIBentryALTinterwordspacing}{\spaceskip=\fontdimen2\font plus
\BIBentryALTinterwordstretchfactor\fontdimen3\font minus \fontdimen4\font\relax}
\providecommand{\BIBforeignlanguage}[2]{{%
\expandafter\ifx\csname l@#1\endcsname\relax
\typeout{** WARNING: IEEEtran.bst: No hyphenation pattern has been}%
\typeout{** loaded for the language `#1'. Using the pattern for}%
\typeout{** the default language instead.}%
\else
\language=\csname l@#1\endcsname
\fi
#2}}
\providecommand{\BIBdecl}{\relax}
\BIBdecl

\bibitem{chen2024dawn}
L.~Chen, Y.~Chen, Z.~Chu, W.~Fang, T.-Y. Ho, Y.~Huang, S.~Khan, M.~Li, X.~Li, Y.~Liang \emph{et~al.}, ``The dawn of ai-native eda: Promises and challenges of large circuit models,'' \emph{arXiv preprint arXiv:2403.07257}, 2024.

\bibitem{li2023ieda}
X.~Li, S.~Tao, Z.~Huang, S.~Chen, Z.~Zeng, L.~Ni, Z.~Huang, C.~Zhuang, H.~Wu, W.~Li \emph{et~al.}, ``ieda: An open-source intelligent physical implementation toolkit and library,'' \emph{arXiv preprint arXiv:2308.01857}, 2023.

\bibitem{ajayi2019openroad}
T.~Ajayi, D.~Blaauw, T.~Chan, C.~Cheng, V.~Chhabria, D.~Choo, M.~Coltella, S.~Dobre, R.~Dreslinski, M.~Foga{\c{c}}a \emph{et~al.}, ``Openroad: Toward a self-driving, open-source digital layout implementation tool chain,'' \emph{Proc. GOMACTECH}, pp. 1105--1110, 2019.

\bibitem{hong2023metagpt}
S.~Hong, X.~Zheng, J.~Chen, Y.~Cheng, J.~Wang, C.~Zhang, Z.~Wang, S.~K.~S. Yau, Z.~Lin, L.~Zhou \emph{et~al.}, ``Metagpt: Meta programming for multi-agent collaborative framework,'' \emph{arXiv preprint arXiv:2308.00352}, 2023.

\bibitem{qian2023communicative}
C.~Qian, X.~Cong, C.~Yang, W.~Chen, Y.~Su, J.~Xu, Z.~Liu, and M.~Sun, ``Communicative agents for software development,'' \emph{arXiv preprint arXiv:2307.07924}, 2023.

\bibitem{wu2023autogen}
Q.~Wu, G.~Bansal, J.~Zhang, Y.~Wu, S.~Zhang, E.~Zhu, B.~Li, L.~Jiang, X.~Zhang, and C.~Wang, ``Autogen: Enabling next-gen llm applications via multi-agent conversation framework,'' \emph{arXiv preprint arXiv:2308.08155}, 2023.

\bibitem{nakajima2023babyagi}
Y.~Nakajima, ``Babyagi,'' \emph{Python. https://github. com/yoheinakajima/babyagi}, 2023.

\bibitem{mandi2023roco}
Z.~Mandi, S.~Jain, and S.~Song, ``Roco: Dialectic multi-robot collaboration with large language models,'' \emph{arXiv preprint arXiv:2307.04738}, 2023.

\bibitem{zhang2023building}
H.~Zhang, W.~Du, J.~Shan, Q.~Zhou, Y.~Du, J.~B. Tenenbaum, T.~Shu, and C.~Gan, ``Building cooperative embodied agents modularly with large language models,'' \emph{arXiv preprint arXiv:2307.02485}, 2023.

\bibitem{park2023generative}
J.~S. Park, J.~O'Brien, C.~J. Cai, M.~R. Morris, P.~Liang, and M.~S. Bernstein, ``Generative agents: Interactive simulacra of human behavior,'' in \emph{Proceedings of the 36th Annual ACM Symposium on User Interface Software and Technology}, 2023, pp. 1--22.

\bibitem{xu2023language}
Z.~Xu, C.~Yu, F.~Fang, Y.~Wang, and Y.~Wu, ``Language agents with reinforcement learning for strategic play in the werewolf game,'' \emph{arXiv preprint arXiv:2310.18940}, 2023.

\bibitem{yao2024hdldebugger}
X.~Yao, H.~Li, T.~H. Chan, W.~Xiao, M.~Yuan, Y.~Huang, L.~Chen, and B.~Yu, ``Hdldebugger: Streamlining hdl debugging with large language models,'' \emph{arXiv preprint arXiv:2403.11671}, 2024.

\bibitem{thakur2023benchmarking}
S.~Thakur, B.~Ahmad, Z.~Fan, H.~Pearce, B.~Tan, R.~Karri, B.~Dolan-Gavitt, and S.~Garg, ``Benchmarking large language models for automated verilog rtl code generation,'' in \emph{2023 Design, Automation \& Test in Europe Conference \& Exhibition (DATE)}.\hskip 1em plus 0.5em minus 0.4em\relax IEEE, 2023, pp. 1--6.

\bibitem{lu2024rtllm}
Y.~Lu, S.~Liu, Q.~Zhang, and Z.~Xie, ``Rtllm: An open-source benchmark for design rtl generation with large language model,'' in \emph{2024 29th Asia and South Pacific Design Automation Conference (ASP-DAC)}.\hskip 1em plus 0.5em minus 0.4em\relax IEEE, 2024, pp. 722--727.

\bibitem{liu2023verilogeval}
M.~Liu, N.~Pinckney, B.~Khailany, and H.~Ren, ``Verilogeval: Evaluating large language models for verilog code generation,'' in \emph{2023 IEEE/ACM International Conference on Computer Aided Design (ICCAD)}.\hskip 1em plus 0.5em minus 0.4em\relax IEEE, 2023, pp. 1--8.

\bibitem{blocklove2023chip}
J.~Blocklove, S.~Garg, R.~Karri, and H.~Pearce, ``Chip-chat: Challenges and opportunities in conversational hardware design,'' \emph{arXiv preprint arXiv:2305.13243}, 2023.

\bibitem{wu2024chateda}
H.~Wu, Z.~He, X.~Zhang, X.~Yao, S.~Zheng, H.~Zheng, and B.~Yu, ``Chateda: A large language model powered autonomous agent for eda,'' \emph{IEEE Transactions on Computer-Aided Design of Integrated Circuits and Systems}, 2024.

\bibitem{liu2023rtlcoder}
S.~Liu, W.~Fang, Y.~Lu, Q.~Zhang, H.~Zhang, and Z.~Xie, ``Rtlcoder: Outperforming gpt-3.5 in design rtl generation with our open-source dataset and lightweight solution,'' \emph{arXiv preprint arXiv:2312.08617}, 2023.

\bibitem{nijkamp2022codegen}
E.~Nijkamp, B.~Pang, H.~Hayashi, L.~Tu, H.~Wang, Y.~Zhou, S.~Savarese, and C.~Xiong, ``Codegen: An open large language model for code with multi-turn program synthesis,'' \emph{arXiv preprint arXiv:2203.13474}, 2022.

\bibitem{huang2024towards}
H.~Huang, Z.~Lin, Z.~Wang, X.~Chen, K.~Ding, and J.~Zhao, ``Towards llm-powered verilog rtl assistant: Self-verification and self-correction,'' \emph{arXiv preprint arXiv:2406.00115}, 2024.

\bibitem{thakur2023autochip}
S.~Thakur, J.~Blocklove, H.~Pearce, B.~Tan, S.~Garg, and R.~Karri, ``Autochip: Automating hdl generation using llm feedback,'' \emph{arXiv preprint arXiv:2311.04887}, 2023.

\bibitem{zhong2023llm4eda}
R.~Zhong, X.~Du, S.~Kai, Z.~Tang, S.~Xu, H.-L. Zhen, J.~Hao, Q.~Xu, M.~Yuan, and J.~Yan, ``Llm4eda: Emerging progress in large language models for electronic design automation,'' \emph{arXiv preprint arXiv:2401.12224}, 2023.

\bibitem{bai2021boom}
C.~Bai, Q.~Sun, J.~Zhai, Y.~Ma, B.~Yu, and M.~D. Wong, ``Boom-explorer: Risc-v boom microarchitecture design space exploration framework,'' in \emph{2021 IEEE/ACM International Conference On Computer Aided Design (ICCAD)}.\hskip 1em plus 0.5em minus 0.4em\relax IEEE, 2021, pp. 1--9.

\bibitem{ma2019cad}
Y.~Ma, Z.~Yu, and B.~Yu, ``Cad tool design space exploration via bayesian optimization,'' in \emph{2019 ACM/IEEE 1st Workshop on Machine Learning for CAD (MLCAD)}.\hskip 1em plus 0.5em minus 0.4em\relax IEEE, 2019, pp. 1--6.

\bibitem{geng2022ppatuner}
H.~Geng, Q.~Xu, T.-Y. Ho, and B.~Yu, ``Ppatuner: Pareto-driven tool parameter auto-tuning in physical design via gaussian process transfer learning,'' in \emph{Proceedings of the 59th ACM/IEEE Design Automation Conference}, 2022, pp. 1237--1242.

\bibitem{geng2022techniques}
H.~Geng, T.~Chen, Q.~Sun, and B.~Yu, ``Techniques for cad tool parameter auto-tuning in physical synthesis: A survey,'' in \emph{2022 27th Asia and South Pacific Design Automation Conference (ASP-DAC)}.\hskip 1em plus 0.5em minus 0.4em\relax IEEE, 2022, pp. 635--640.

\bibitem{geng2022ptpt}
H.~Geng, T.~Chen, Y.~Ma, B.~Zhu, and B.~Yu, ``Ptpt: Physical design tool parameter tuning via multi-objective bayesian optimization,'' \emph{IEEE transactions on computer-aided design of integrated circuits and systems}, vol.~42, no.~1, pp. 178--189, 2022.

\bibitem{zhai2023microarchitecture}
J.~Zhai and Y.~Cai, ``Microarchitecture design space exploration via pareto-driven active learning,'' \emph{IEEE Transactions on Very Large Scale Integration (VLSI) Systems}, 2023.

\bibitem{jung2021metrics2}
J.~Jung, A.~B. Kahng, S.~Kim, and R.~Varadarajan, ``Metrics2. 1 and flow tuning in the ieee ceda robust design flow and openroad iccad special session paper,'' in \emph{2021 IEEE/ACM International Conference On Computer Aided Design (ICCAD)}.\hskip 1em plus 0.5em minus 0.4em\relax IEEE, 2021, pp. 1--9.

\bibitem{nardi2019practical}
L.~Nardi, D.~Koeplinger, and K.~Olukotun, ``Practical design space exploration,'' in \emph{2019 IEEE 27th International Symposium on Modeling, Analysis, and Simulation of Computer and Telecommunication Systems (MASCOTS)}.\hskip 1em plus 0.5em minus 0.4em\relax IEEE, 2019, pp. 347--358.

\bibitem{hosny2021characterizing}
A.~Hosny and S.~Reda, ``Characterizing and optimizing eda flows for the cloud,'' \emph{IEEE Transactions on Computer-Aided Design of Integrated Circuits and Systems}, vol.~41, no.~9, pp. 3040--3051, 2021.

\bibitem{kellerer2004multiple}
H.~Kellerer, U.~Pferschy, D.~Pisinger, H.~Kellerer, U.~Pferschy, and D.~Pisinger, ``The multiple-choice knapsack problem,'' \emph{Knapsack Problems}, pp. 317--347, 2004.

\bibitem{dudzinski1987exact}
K.~Dudzi{\'n}ski and S.~Walukiewicz, ``Exact methods for the knapsack problem and its generalizations,'' \emph{European Journal of Operational Research}, vol.~28, no.~1, pp. 3--21, 1987.

\bibitem{opencores}
``{Opencores},'' [Online]. Available: \url{https://opencores.org/}.

\end{thebibliography}

\end{document}